\documentclass[twocolumn,prl,a4paper]{revtex4}
\usepackage{amsthm}
\usepackage{bbm}
\usepackage{amssymb}
\usepackage{amsmath}

\newtheorem{thm}{Theorem}
\newtheorem*{thm*}{Theorem}

\theoremstyle{definition}

\theoremstyle{remark}

\newcommand{\lra}{\longrightarrow}

\newcommand{\N}{\mathbb{N}}
\newcommand{\K}{\mathbb{K}}
\newcommand{\R}{\mathbb{R}}
\newcommand{\C}{\mathbb{C}}

\newcommand{\Q}{\mathcal{Q}}

\newcommand{\1}{\mathbbm{1}}

\renewcommand{\>}{\rangle}

 \DeclareMathOperator{\tr}{tr}

\begin{document}

\title{Operator Space theory: a natural framework for Bell inequalities}

\author{M. Junge$^1$}
\author{C. Palazuelos$^2$}
\author{D. P\'{e}rez-Garc\'{\i}a$^2$}
\author{I. Villanueva$^2$}
\author{M.M. Wolf$^3$}
 \affiliation{
  $^1$Department of Mathematics, University of Illinois at Urbana-Champaign, Illinois 61801-2975, USA\\
  $^2$Departamento Analisis Matematico and IMI, Universidad Complutense de Madrid, 28040 Madrid, Spain\\
  $^3$Niels Bohr Institute, Blegdamsvej 17, 2100 Copenhagen, Denmark}

\begin{abstract}
In this letter we show that the field of Operator Space Theory provides a general and powerful mathematical framework for arbitrary Bell inequalities, in particular regarding the scaling of their violation within quantum mechanics. We illustrate the power of this connection by showing that bipartite quantum states with local Hilbert space dimension $n$ can violate a Bell inequality by a factor of order $\frac{\sqrt{n}}{\log^2n}$ when observables with $n$ possible outcomes are used. Applications to resistance to noise, Hilbert space dimension estimates and communication complexity are given.
\end{abstract}


\maketitle

Bell inequalities \cite{WernerWolf} were originally proposed by Bell \cite{Bell} in 1964 as
a way of testing the validity of Einstein-Podolski-Rosen's believe that local hidden
variable models are a possible underlying explanation of physical reality \cite{EPR}. Nowadays, they are at the heart of the modern development of
Quantum Information, with applications in a wide variety of areas: quantum key
distribution \cite{QKD, Acin}, entanglement detection, multipartite interactive proof systems \cite{MIP,KRT}, communication
complexity \cite{Buhrman}, Hilbert space
dimension estimation \cite{dimension,dimension2,Brunner,PWJPV},
etc.

Despite their importance, very few is known beyond very particular cases and examples. One reason for that is that so far there was no suitable mathematical tool for them.  In \cite{PWJPV}, we showed how for the special case of {\it correlation} Bell inequalities, \emph{Operator Space Theory}---a modern field in mathematical analysis---provides exactly the right language and tools to tackle some of the more difficult problems. There we used operator space techniques to solve an old question of Tsirelson \cite{Tsirelson}: the existence of unbounded violations for tripartite correlation Bell inequalities. At the same time this established a new result in (formerly) pure mathematics about the generalization of a celebrated result by Grothendieck. Following and extending these lines, we are now able to show that Operator Space Theory is indeed the right mathematical theory to deal with {\it arbitrary} Bell inequalities, not restricted to the correlation case based on two-outcome measurements.

The aim of the present paper is to sketch the deep relation between the field of Operator Space Theory on the one hand and quantum mechanical Bell inequality violations on the other. Once this connection is established, the language of Operator Spaces allows to derive various new results and considerably strengthen known ones. The mathematical part of these derivations  goes beyond the scope of the present paper and is presented elsewhere \cite{CMP2}.

We  illustrate the power of the new methods by showing that quantum mechanics allows for
 violations of bipartite Bell inequalities of the order $\frac{\sqrt{n}}{\log^2n}$ when given $n$ dimensional
Hilbert spaces and observables with $n$ possible measurement outcomes. This result in turn implies better
Hilbert space dimension witnesses and non-local quantum distributions
with better resistance to noise---something desirable on the way to
loophole free Bell tests. We also discuss  implications for quantum communication complexity theory.

 \subsection{Bell Inequalities}

We consider a setup where two distant observers, Alice and Bob, each receives one part of a bipartite correlated system and repeatedly performs one out of several measurements.
Assume that both can choose among $N$ different observables which are labeled by $x$ for Alice and $y$ for Bob, respectively, and let $a,b=1,\ldots, M$ be the possible outcomes of their measurements.  We denote by $p(a,b  |x,y)$
the probability that Alice and Bob get outcomes $a,b$ when performing
measurements $x,y$. Note that
we can consider a distribution $P=\{p(a,b|x,y)\}$ as an element of the space ${\cal R}=\mathbb R^{M^2
N^2}$.

 A distribution $P$ admits a local hidden variable (LHV) representation if it is of the form
 \begin{equation}\label{eq:LHV}
 p(a,b|x,y)=\int_\Lambda q(a|x,\lambda)q'(b|y,\lambda) \mu(d\lambda),\end{equation} where $\Lambda$ is a possible ``hidden variable''  space and $q,q'$ and $\mu$ are probability distributions.
We use the
notation $\mathcal L$ for the set of all LHV distributions (for given $N$ and $M$) and
$\mathcal Q$ for the set of distributions which can arise from quantum mechanics. That is $p(ab|xy)\in \Q$ if there exist POVMs $\{E_a^x\}$ for Alice and $\{F_b^y\}$ for Bob and a quantum state $\rho_{AB}$ such that \begin{equation}\label{eq:quantum}p(ab|xy)=\tr\left(\rho_{AB} E_a^x\otimes F_b^y\right).\end{equation} $\Q_d$  will denote $\Q$
with the extra restriction that the local Hilbert spaces of both Alice and Bob
 are $d$-dimensional,

It is well known that $\mathcal L\subsetneqq \mathcal Q\subset {\cal R}$.

Since $\mathcal L$ is a polytope it can be characterized by a finite set of linear inequalities---
{\em Bell inequalities}. In general we will assign a Bell inequality to every linear functional $T$ in the dual of $\cal R$.
The corresponding Bell inequality is then a
statement of the form
$$\mbox{For every }P\in \mathcal L, \, \left|\sum_{a,b,x,y}T_{x,y}^{a,b}\; p(a,b\,|x,y)\right|\leq C.$$

We can shorten the above
notation by writing $|\langle T,P\rangle|\leq C$ and we will simply refer to the functional $T$ as a Bell inequality, assuming that $C$
 is defined by $C=\sup_{P\in \mathcal L}|\langle T,P\rangle|$.

Since $\mathcal L\subsetneqq \mathcal Q$, Quantum Mechanics allows for a violation of at least some of these inequalities. Quantitatively, we define the
{\em violation of a Bell inequality $T$ by a distribution $Q$} as
 \begin{equation}\label{eq:viol-1}\frac{|\langle T,Q\rangle|}{\sup_{P\in \mathcal
L}|\langle T,P\rangle|}.\end{equation} Taking the ratio and the absolute value is crucial for a meaningful definition of the
{\it amount of violation}. If instead of the ratio, one takes for instance the difference, a change of scale $T\rightarrow\lambda T$ would lead to arbitrary violations. If one removes instead the absolute value, the same happens via an offset.

As we will see below, the amount of violation defined by (\ref{eq:viol-1}) exhibits some resource character and has a clear physical interpretation in terms of noise resistance.  We will be interested in the {\em maximum violation}
$$\nu(Q)=\sup_T \frac{|\langle T,Q\rangle|}{\sup_{P\in \mathcal L}|\langle
T,P\rangle|}.$$

It is sometimes convenient to consider \emph{incomplete} probability distributions (summing up to less than one), which are the ones obtained when Alice and Bob carry out incomplete measurements. They are characterized then also by equations (\ref{eq:LHV}) and (\ref{eq:quantum}) when changing the equalities $\sum_{a}q(a|x,\lambda) = 1$, $\sum_{b}q'(b|y,\lambda) = 1$,  $\sum
_aE_x^a= \1$, $\sum
_bF_y^b= \1$ to inequalities. We will denote these sets by $\mathcal L^{in}$ and $\mathcal Q^{in}$. By adding/removing one output, it is easy to see that
\begin{equation}\label{eq:aqfe}\sup_{Q\in\mathcal{Q}_n}\nu(Q)\ge\sup_T \frac{\sup_{Q \in
\mathcal{Q}^{in}_n}|\langle T,Q\rangle|}{\sup_{ P\in \mathcal{L}^{in}}|\langle T,P\rangle|}.\end{equation}

\subsection{The mathematical language: Operator Spaces}

We will now sketch  how the right hand side of Eq.(\ref{eq:aqfe}) can be seen as the quotient of two natural norms arising in operator space theory.
The mathematical theory of Operator Spaces started developing in the late 80's, but it already has offered powerful new tools for the
solution of long standing mathematical problems (see \cite{Pisierbook} and the references therein).  Essentially, an
operator space is a complex Banach space $E$ together with a sequence of
``reasonable'' norms in the spaces $M_n\otimes E=M_n(E)$, where
$M_n$ is the space of square matrices of order $n$ endowed with the operator norm and $M_n(E)$ is
the space of square matrices of order $n$ with entries in $E$. This turns out to be equivalent to consider $E$ as
a closed subspace of $B(H)$ (fixing the embedding) and defining the
norm in $M_n(E)$ as the norm inherited by the embedding
$M_n(E)\subset M_n(B(H))\approx B(\ell_2^n(H))$. $B(H)$ denotes here the space of bounded linear operators on a Hilbert space $H$ and, as made explicit below, $\ell_2^n$ is the $n$-dimensional Hilbert space.

In Banach space theory, the morphisms are the linear and bounded maps. A linear map $u:E\lra F$ is bounded if its norm $\|u\|=\sup_{\|x\|_E\le 1}\|u(x)\|_F$ is finite. The same happens for bilinear maps, $T:E\times F\lra G$, where now the norm is $\|T\|=\sup_{\|x\|_E\le 1, \|y\|_F\le 1}\|T(x,y)\|_G$. In the case of operator spaces, the relevant morphisms to capture the extra structure are the {\em
completely bounded mappings}. Given a linear map between operator spaces $u:E\lra F$, $u$ is completely bounded if $\|u\|_{cb}:=\sup_n\|\1_n \otimes u: M_n(E)\lra M_n(F)\|<\infty$. In that case, $\|u\|_{cb}$ is the completely bounded norm of $u$.
Given two operator spaces $E,F$, a bilinear form $T:E\times F
\longrightarrow \mathbb C$ is {\em completely bounded} \footnote{Some references talk about {\it jointly} completely bounded bilinear maps, since there was a different previous definition of completely bounded bilinear maps associated to the Haagerup tensor norm} if all the
induced bilinear forms $T_n:=\1_n\otimes \1_n\otimes T:M_n(E)\times M_n(F) \longrightarrow M_n\otimes M_n\otimes \mathbb C\approx M_{n^2}$
are uniformly bounded in the Banach space sense. In that case, we
define its {\em completely bounded norm} by $\|T\|_{cb}=\sup_{n\in
\mathbb N} \|T_n\|$.

Given a Banach space $E$, one defines its dual  $E^*$ as the space of bounded linear maps from $E$ to $\K$ with the norm defined above. $\K$ here is the scalar field which can be the real $\R$ or the complex $\C$ numbers. So for instance, for the space $\ell_p^N$, defined as $\K^N$ with the norm $\|x\|_p=\left(\sum_{i=1}^N|x_i|^p\right)^\frac{1}{p}$ for $1\le p<\infty$ and $\|x\|_\infty=\max_i|x_i|$, we have that $(\ell_p^{N})^*=\ell_q^N$ for $\frac{1}{p}+\frac{1}{q}=1$. In particular $(\ell_\infty^{N})^*=\ell_1^N$. We can also consider the space  $\ell_\infty^N(\ell_1^M)$, defined as the space $\K^N\otimes \K^M$ endowed with the norm $\|x\|=\max_{i=1}^N\sum_{j=1}^M|x_{i,j}|$.

The same can be done for operator spaces. The way to define $M_n(E^*)$ is simply by identifying $M_n(E^*)=CB(E,M_n)$. With this construction, we can start to define {\it natural} operator space structures in some Banach spaces. The starting point is the space $\ell_\infty^N$. It can be mapped trivially into the diagonal of $M_N=B(\ell_2^N)$ and in this way it acquires the operator space structure given by the norms $\|\sum_{i} A_i \otimes e_i\|_{M_n(\ell_\infty^N)}=\max_i \|A_i\|_{M_n}$, where $\{e_i\}$ denotes the canonical basis of $\C^N$. By duality, this allows to define then an operator space structure in $\ell_1^N$. But also in $\ell_\infty^N(\ell_1^M)$, by identifying $M_n(\ell_\infty^N(\ell_1^M))=CB(\ell_1^N,M_n(\ell_1^M))$.

Let us see now the connection with Bell inequalities. Let us take a Bell inequality $T$ and compute its norm as a {\it real} bilinear form $T:\ell_\infty^N(\ell_1^M)\times \ell_\infty^N(\ell_1^M)\lra \R$. It gives
$$\sup\{\left|\sum_{a,b,x,y}T_{x,y}^{a,b} p_{a,x}q_{b,y}\right|: \max_x\sum_a|p_{a,x}|,\max_y\sum_b|p_{b,y}|\le 1 \}.$$
If $p_{a,x}$ were positive, one would have $\max_x\sum_ap_{a,x}\le 1$ which allows to identify $p_{a,x}$ with a local incomplete distribution $p(a|x)$ for Alice---the same for Bob. Then, decomposing both $p_{a,x}$ and $q_{b,y}$ in positive and negative part one gets
$$\sup_{P\in\mathcal{L}^{in}}|\langle T,P\rangle |\leq \|T\|\leq 4\sup_{P\in\mathcal{L}^{in}}|\langle
T,P\rangle|.$$
If we compute now the norm of $T_n$ as a {\it complex} bilinear map, it gives
$$\sup\left|\sum_{a,b,x,y}T_{x,y}^{a,b} \tr(E_a^x\otimes F_b^y\rho_{AB})\right|,$$ where the sup is restricted to $\tr{|\rho_{AB}|}= 1$, $\|\sum_{a,x} E_a^x \otimes e_x\otimes e_a\|_{M_n(\ell_\infty^N(\ell_1^M))} \le 1$, $\|\sum_{b,y} F_b^y\otimes e_y\otimes e_b\|_{M_n(\ell_\infty^N(\ell_1^M))}\le 1$. With some operator space tools (see \cite{CMP2}), one can obtain that, if the matrices $E_a^x\ge 0$, then $\|\sum_{a,x} E_a^x \otimes e_x\otimes e_a\|_{M_n(\ell_\infty^N(\ell_1^M))}\le 1$ if and only if $\sum_a E_a^x\le \1$ for all $x$, and we get an incomplete POVM. In this case we can also assume $\rho_{AB}$ to be positive and hence a proper quantum density matrix. In contrast to the Banach space situation, now the required decomposition in terms of positive matrices is not trivial, but relies on Wittstock's factorization Theorem \cite{Pisierbook} (see \cite{CMP2} for details). This finally gives
$$\sup_{Q\in \mathcal Q_n^{in}}|\langle T,Q\rangle|\leq \|T_n\| \leq 16 \sup_{Q\in \mathcal Q_n^{in}}|\langle
T,Q\rangle|, $$
and we get that, for a given Bell inequality $T$,$$\frac{1}{16}\frac{\|T_n\|}{\|T\|}\le\frac{\sup_{Q\in \mathcal Q_n ^{in}}|\langle T,Q\rangle|}{\sup_{P\in\mathcal L^{in}}|\langle
T,P\rangle|}\leq 4\frac{\|T_n\|}{\|T\|}.$$

Due to the appearing constants it may be difficult to exactly determine a specific Bell inequality violation. However, since the constants are universal and in particular independent of the Hilbert space dimension $n$, the above relation enables us to determine the scaling of the maximal violation with increasing $n$. For these statements we will use the Landau symbols $\Omega$ and $O$ denoting asymptotic lower and upper bounds, respectively.

\subsection{The applications: Unbounded violations}

Using Operator Space Theory we can now  estimate $\sup_T\frac{\|T_n\|}{\|T\|}$ (see \cite{CMP2} for details). Translated into the language of Bell inequalities the obtained result is the following.

\begin{thm}\label{main}
For every $n\in \N$ there exists a bipartite quantum probability
distribution $Q\in\mathcal Q$ obtained from $N=[2^\frac{\log^2n}{2}]^n$ observables with $M=n+1$
outcomes and acting on Hilbert spaces of local dimension $n$ such that
$$\nu(Q)=\Omega\left(  \frac{\sqrt{n}}{\log^2n}\right).$$
\end{thm}
In \cite{CMP2} we also show that the scaling with respect to the Hilbert space dimension is not too far from optimal by proving the upper bound $ \nu(Q)={\rm O}(n) $ for the case of Hilbert spaces with dimension $n$ and arbitrary numbers of observables and outcomes. Similarly, if only the number of measurement outcomes is fixed to $n$ then $\nu(Q)={\rm O}(n^2)$ for any number of observables and any Hilbert space dimension \cite{DKLR}.

\subsubsection{Prior bipartite unbounded violations}

As pointed out by Tsirelson \cite{Tsirelson}, Grothendieck's
Theorem shows that we can not obtain unbounded violations in the case of
correlation matrices for bipartite systems---we have to make use of the full probability distribution.

The first unbounded violations of bipartite Bell inequalities were obtained as an application of Raz parallel repetition theorem \cite{Raz}, which
 ensures that the parallel repetition of the magic square
game provides a violation $\Omega(n^x)$ for some $x>0$ with $n$ inputs, $n$ outputs
and a Hilbert space of dimension $n$. The
best available bounds using these techniques
seems to be not much better than $\Omega(n^{10^{-5}})$.

In \cite{KRT}, the authors used a deep result of Khot and Vishnoi in the context of complexity
theory \cite{KV} to get violations of order $\Omega(n^{\frac{1}{54}})$ with $n$ outputs and $\frac{2^n}{n}$ inputs.

With this in mind, our $\Omega\left(\frac{\sqrt{n}}{\log^2n}\right)$ violation
with $n$ outputs and Hilbert space dimension $n$ can be seen as a considerable improvement over the previous results. One can not rule out, however, the possibility of obtaining similar bounds to ours pushing forward the previous kind of techniques.

\subsubsection{Resistance to noise}\label{sec:noise}

 To emphasize the relevance of the chosen definition for `maximal violation' we provide an operational meaning for $\nu(Q)$: it quantifies the amount of  noise that a given quantum distribution $Q$ can withstand before admitting an explanation within a local hidden variable model. Here we allow for any type of noise which itself admits a LHV description. More precisely, we  show that  $\nu(Q)=\frac{2}{\pi(Q)}-1$, where $$\pi(Q)=\inf\{\pi: \pi Q+(1-\pi)P\not\in \mathcal{L} \text{  for all  }P\in \mathcal{L}\}.$$
A particular type of  noise is for instance the one given by imperfect  detectors, which has been widely discussed in relation to the ``detection loophole'' (see
for instance
\cite{Buhrman,inefficient}).

Our main result proves the existence of quantum
probability distributions with $n$ outputs and Hilbert spaces of
dimension $n$ which can withstand {\it any} LHV noise with relative
strength $O(1-\frac{\log^2(n)}{\sqrt{n}})$. Note \cite{CMP2} that $O(1-\frac{1}{n})$ corresponds to
an upper bound for the maximal possible resistance to noise.

\subsubsection{Bounds for the Hilbert space dimension}

Motivated by the role it plays for quantum cryptography \cite{Acin}, several protocols for estimating the Hilbert
space dimension have been proposed \cite{dimension,dimension2,Brunner}.  A ``dimension witness'' \cite{Brunner} for Hilbert space dimension
$d$ is simply a ``Bell-type inequality'' $M$ such that
$|\<M,P\>|\le C$ for all $P\in \Q_d$, and for which
there exist $P'\in \Q_n$, $n>d$ with $|\<M,P'\>|> C$. For measurements with binary outcomes it is shown in \cite{dimension} how to get dimension witnesses for any dimension, with the drawback that the resolution of the considered witnesses is bounded by Grothendieck's constant $K_G$ and indeed could vanish with increasing dimension. In particular the violation of a bipartite correlation Bell inequality itself can clearly not accurately separate between different Hilbert space dimensions since it is always (for every $d$) constrained between $\sqrt{2}$ and Grothendieck's $K_G\leq 1.783$.

Our results show that this changes if one allows  either for more parties or for measurements with more outcomes. In \cite{PWJPV} we showed that a violation of a tripartite Bell inequality of order $\sqrt{d}$ certifies that the smallest involved Hilbert space dimension is at least $d$. Similarly, the presently obtained upper bound shows that a violation of a bipartite inequality of order $d$ requires a Hilbert space of dimension at least $d$. The obtained lower bound implies that such a violation is indeed possible if the actual dimension is of order $n$ with $\sqrt{n}/\log^2(n)\geq d$.

\subsubsection{Communication complexity}

A basic problem in communication complexity is the following \cite{Buhrman}:
Assume Alice is given an input $x$ and must give an output $a$ whereas Bob is
given an input $y$ and must give an output $b$. How many bits do they have to communicate in order to reproduce a
given probability distribution $P=p(ab|xy)$?.

In \cite{ReTo} the authors show that only $2$ bits of communication
suffice to simulate classically all possible quantum correlations
matrices. That is, in this case entanglement only saves $2$
bits. On the contrary, Gavinsky \cite{Gavinsky}, culminating a
remarkable series of exponential quantum-classical savings
\cite{complexity} has shown
recently that there exist quantum distributions $P$ of inputs of
length (number of bits) $n$, outputs of length ${\rm
O}(\log^2n\log\log n)$ and obtained with ${\rm O}(\log^2n\log\log
n)$ EPR pairs that cannot be simulated classically even if we allow
for small errors and ${\rm O}\left( \frac{n^{1/4}}{\log^2n}\right)$
bits of two-way communication.

In \cite{DKLR} it is shown how, for a given quantum probability
distribution $P\in \Q$, $\log(\nu(P))$ is a lower bound to the
number of bits needed to be transmitted between Alice and Bob in
order to simulate $P$ classically. Our main Theorem \ref{main}
shows then new examples of quantum  distributions $P\in Q$ with $n$
outputs $\exp(n)$ inputs and Hilbert space dimension $n$ which need
the transmission of ${\rm \Omega}(\log(n))$ bits in order to be simulated
classically. The saving in the communication complexity obtained in
this way is, however, exponentially worse than the one in the
example of Gavinsky \cite{Gavinsky}.

\section{Conclusion}

We have seen that the theory of Operator Spaces not only provides a perfect framework
to formulate quantum violations of Bell inequalities but also provides new mathematical tools for giving
upper and lower estimates to the order of magnitude of the violation. In particular, we have obtained nearly optimal quantum Bell violations in the bipartite case. We strongly expect this connection to be exploited further in the future, giving new insights in most areas related to Bell inequalities: quantum cryptography, multipartite interactive proof systems, communication complexity, etc.

The authors are grateful to the organizers of the \emph{Operator Structures in Quantum Information Workshop}, held in Toronto during July 6-10, 2009; where part of this work was developed. M. Junge is partially supported by the NSF grant DMS-0901457. C. Palazuelos, D. Perez-Garcia and I. Villanueva are partially supported by Spanish grants I-MATH, MTM2008-01366 and CCG08-UCM/ESP-4394. M.M.
Wolf acknowledges support by QUANTOP and the Danish Natural Science
Research Council(FNU).

\end{document}